\documentclass[twocolumn,showpacs,preprintnumbers,amsmath,amssymb]{revtex4}

\usepackage{graphicx}% Include figure files
\usepackage{dcolumn}% Align table columns on decimal point
\usepackage{bm}% bold math

\begin{document}

\preprint{DF/IST-07.2004}

\title{Global embedding of $\bm D$-dimensional black holes with a
cosmological constant in Minkowskian spacetimes: Matching between
Hawking temperature and Unruh temperature}

\author{Nuno Loureiro Santos}
\email{nlsantos@fisica.ist.utl.pt} \affiliation{ Centro
Multidisciplinar de Astrof\'{\i}sica - CENTRA, Departamento de
F\'{\i}sica, Instituto Superior T\'ecnico, Universidade T\'ecnica
de Lisboa, Av. Rovisco Pais 1, 1049-001 Lisboa, Portugal}
\author{\'Oscar J. C. Dias}
\email{odias@perimeterinstitute.ca} \affiliation{ Department of
Physics, University of Waterloo, Waterloo, Ontario N2L 3G1, Canada\\
and\\
Perimeter Institute for Theoretical Physics, 31 Caroline St. N.,
Waterloo, Ontario N2L 2Y5, Canada }
\author{Jos\'e P. S. Lemos}
\email{lemos@fisica.ist.utl.pt} \affiliation{ Centro
Multidisciplinar de Astrof\'{\i}sica - CENTRA, Departamento de
F\'{\i}sica, Instituto Superior T\'ecnico, Universidade T\'ecnica
de Lisboa, Av. Rovisco Pais 1, 1049-001 Lisboa, {}Portugal}

%\date{\today}

\begin{abstract}
We study the matching between the Hawking temperature of a large
class of static $D$-dimensional black holes and the Unruh
temperature of the corresponding higher dimensional Rindler
spacetime.  In order to accomplish this task we find the global
embedding of the $D$-dimensional black holes into a higher
dimensional Minkowskian spacetime, called the global embedding
Minkowskian spacetime procedure (GEMS procedure).  These global
embedding transformations are important on their own, since they
provide a powerful tool that simplifies the study of black hole
physics by working instead, but equivalently, in an accelerated
Rindler frame in a flat background geometry. We discuss neutral
and charged Tangherlini black holes with and without cosmological
constant, and in the negative cosmological constant case, we
consider the three allowed topologies for the horizons (spherical,
cylindrical/toroidal and hyperbolic).
\end{abstract}



\maketitle
%%%%%%%%%%%%%%%%%%%%%%%%%%%%%%%%%%%%%%%%%%%%%%%%%%%%%%%%%%%%%%%%
\section{\label{sec:intro}Introduction}
%%%%%%%%%%%%%%%%%%%%%%%%%%%%%%%%%%%%%%%%%%%%%%%%%%%%%%%%%%%%%%%%

In \cite{hawking} Hawking has shown that black holes are not black
at all: they emit radiation, through quantum effects, with a
characteristic thermal spectrum of a blackbody with temperature
$T_{\rm H}$ proportional to the horizon surface gravity $k_{\rm
h}$. More concretely, an observer at rest at a constant radial
coordinate close to the event horizon measures a Hawking
temperature given by (we use units in which the light velocity,
the Boltzmann constant and the Planck constant are set equal to
one)
\begin{equation}
\label{hawk} T_{\rm H}=\frac{1}{2\pi} \frac{k_{\rm
h}}{\sqrt{-g_{00}}}\,,
\end{equation}
where $g_{00}$ is the time-time component of the gravitational
metric. Later on, Unruh \cite{unruh} has remarked that the
detection of thermal radiation is not restricted to an observer in
the vicinity of a black hole horizon. Indeed, even in flat
spacetime, an observer that moves with a constant acceleration
$a$, a Rindler observer, will encounter an acceleration horizon
and will also detect thermal radiation. The associated Unruh
temperature $T_{\rm U}$ is given by
\begin{equation}
\label{unruh} T_{\rm U}=\frac{a}{2\pi}.
\end{equation}

Once the Hawking effect is known to exist, one can argue that the
Unruh effect follows from the equivalence principle. Indeed,
according to this principle, the effects measured by an observer
that is at rest in the vicinity of a black hole horizon are the
same as the effects measured by an uniformly accelerated observer
in a flat spacetime. Moreover, the equivalence principle also
suggests that the connection (defined by Eq.(\ref{hawk})) between
the temperature and the surface gravity also holds for Rindler
motions, with $k_{\rm h}$ being now the surface gravity of the
acceleration horizon \cite{waldbook}.

The strong connection between the Hawking and the Unruh
temperatures appear in a different context, when one embeds a
lower dimensional curved spacetime containing a black hole into
higher dimensional flat spacetime with one or more timelike
coordinates.

That such an embedding can be done for any 4-dimensional black
hole geometry was shown in \cite{goenner}.  Historically, the
first embedding of the Schwarzschild black hole in a 6-dimensional
flat spacetime was performed by Kasner \cite{kasner} (he himself
noted that an embedding in 5 dimensions is impossible
\cite{kasner2}). A vast list of embedding transformations for
other 4-dimensional black hole geometries are presented in
\cite{rosen}. All these embeddings have the drawback that they are
incomplete, since they cannot be extended past the event horizon,
for radius less than the black hole radius. The complete, or
global, embedding of the Schwarzschild black hole in a
6-dimensional flat spacetime was presented in \cite{fronsdal}, for
a review see \cite{goenner}.  Applying this procedure
appropriately, one can then map the black hole horizon to its
acceleration horizon counterpart in the higher dimensional flat
spacetime. The global embedding of a black hole geometry into a
higher dimensional Minkowskian spacetime is generically called
global embedding Minkowskian spacetime procedure, or GEMS
procedure, for short. Note that any flat spacetime with one or
more time coordinates is been called here a Minkowski spacetime.
The GEMS procedure is an important procedure on its own, since it
provides a powerful tool that simplifies the study of black hole
physics by working instead, but equivalently, in a flat background
geometry.

Through the GEMS procedure one can match the Hawking temperature
associated with the black hole horizon with the Unruh temperature
associated with the acceleration horizon of the higher dimensional
Rindler spacetime.  This matching has been first confirmed in the
de Sitter (dS) geometry \cite{birrel,narnhofer}, and in the
anti-de Sitter (AdS) geometry \cite{deser1}, in 4-dimensional
spacetimes. It was in \cite{deser1} that the name GEMS appeared.
Soon after, this unified computation of the temperatures has been
verified for the Schwarzschild black hole \cite{deser2}, for the
Reissner-Nordstr\"{o}m black hole \cite{deser3}, and for the
Schwarzschild and Reissner-Nordstr\"{o}m black holes in a
asymptotically dS and AdS 4-dimensional spacetimes
\cite{deser3,kps}, as well as for the 3-dimensional
Ba\~nados-Teilelboim-Zanelli black hole \cite{deser3}.  When
dealing with temperature issues, it is essential that one works
with a global embedding as an incomplete embedding leads to
observers for whom there is no horizon and no temperature
\cite{deser3}.

In this paper we will study the connection between the Hawking
effect and the Unruh effect for the static black hole solutions in
the higher dimensional Einstein-Maxwell theory, generalizing thus
the 4-dimensional results of Deser and Levin \cite{deser3}. The
higher dimensional counterparts of the Schwarzschild and of the
Reissner-Nordstr\"{o}m black holes $-$ the Tangherlini black holes
$-$ have been found and discussed in \cite{tangherlini}. The
higher dimensional Schwarzschild and Reissner-Nordstr\"{o}m black
holes in an asymptotically de Sitter (dS) spacetime and in an
asymptotically anti-de Sitter (AdS) spacetime have also been
discussed in \cite{tangherlini}. Now, in an asymptotically AdS
4-dimensional background, besides the black holes with spherical
topology, there are also solutions with planar, cylindrical or
toroidal topology found and discussed in \cite{Toroidal_4D}, and
black holes with hyperbolic topology analyzed in
\cite{topological}. The higher dimensional extensions of these
non-spherical AdS black holes are already known. Namely, the
$D$-dimensional AdS black holes with planar, cylindrical or
toroidal topology were discussed in \cite{Birmingham-TopDdim}, and
the $D$-dimensional AdS black holes with hyperbolic topology were
analyzed in  \cite{Birmingham-TopDdim,DehghaniRotTopol}. We will
find the global embedding transformations that describe all the
above $D$-dimensional black holes in a higher dimensional flat
background, and then we use them to study the matching between the
Hawking and the Unruh temperatures.

In section \ref{sec:outline} we outline the GEMS procedure. Then,
the black holes in an asymptotically AdS backgrounds will be
analyzed in Section \ref{sub:menor} and the dS background in
\ref{sub:maior}, while the black holes in an asymptotically flat
background will be discussed in Section \ref{sub:flat}.

%%%%%%%%%%%%%%%%%%%%%%%%%%%%%%%%%%%%%%%%%%%%%%%%%%%%%%%%%%%%%%%%%%%%
\section{\label{sec:outline}Black hole geometries. Outline of the
GEMS procedure}
%%%%%%%%%%%%%%%%%%%%%%%%%%%%%%%%%%%%%%%%%%%%%%%%%%%%%%%%%%%%%%%%%%%%

%%%%%%%%%%%%%%%%%%%%%%%%%%%%%%%%%%%%%%%%%%%%%%%%%%%%%%%%%%%%%%%%%%%%
\subsection{\label{sec:BH Ddim}Static black holes in higher dimensions}
%%%%%%%%%%%%%%%%%%%%%%%%%%%%%%%%%%%%%%%%%%%%%%%%%%%%%%%%%%%%%%%%%%%%
In a higher dimensional background with a cosmological constant
$\Lambda$, the Einstein-Maxwell equation is
\begin{equation}
\label{EM} R_{\mu\nu}-\frac{1}{2}{R}g_{\mu\nu}+
\frac{(D-1)(D-2)}{6}\Lambda g_{\mu\nu}=8\pi {T}_{\mu\nu}
\end{equation}
where ${R}_{\mu\nu}$ is the Ricci tensor and ${T}_{\mu\nu}$ is the
electromagnetic energy-momentum tensor (for the corresponding
action see \cite{DdimBH}). This equation allows a three-family of
static black hole solutions, parameterized by the constant $k$
which can take the values $1,0,-1$, and whose gravitational field
is described by
\begin{eqnarray}
d s^2 = - f(r)\, dt^2 + f(r)^{-1}\,dr^2+r^2 (d\,\Omega_{D-2}^k)^2,
\label{gen metric}
\end{eqnarray}
where
\begin{eqnarray}
f(r) =
k-\frac{\Lambda}{3}\,r^2-\frac{M}{r^{D-3}}+\frac{Q^2}{r^{2(D-3)}}\:,
\label{gen f}
\end{eqnarray}
$D\geq 4$, and the mass parameter $M$ and the charge parameter $Q$
are proportional to the Arnowitt-Deser-Misner mass and electric
charge, respectively \cite{myersperry}.
In Eq. (\ref{EM}) the coefficient of the $\Lambda$ term was choosen
such that $f(r)=k-(\Lambda /3)r^2$ when $M=0$ and $Q=0$, as occurs
in $D=4$. For $k=1$, $k=0$ and $k=-1$ one has, respectively,
\begin{eqnarray}
\label{angular} (d\Omega_{D-2}^k)^2 \!\!\!&=& \!\!\!
d\theta_1^2+\sin^2\theta_1\,d\theta_2^2+ \cdots
+\prod_{i=1}^{D-3}\sin^2\theta_i\,d\theta_{D-2}^2\,, \nonumber \\
(d\Omega_{D-2}^k)^2 \!\!\!&=& \!\!\! d\theta_1^2+d\theta_2^2+
d\theta_3^2+\cdots +d\theta_{D-2}^2\,, \nonumber \\
(d\Omega_{D-2}^k)^2 \!\!\!&=& \!\!\!
d\theta_1^2+\sinh^2\theta_1\,d\theta_2^2+\! \cdots \\
\!&+&\sinh^{2}\theta_1\!\!\prod_{i=2}^{D-3}\sin^2\theta_i\,
d\theta_{D-2}^2\,. \nonumber
\end{eqnarray}
The family with $k=1$ describes the Tangherlini black holes with
spherical topology \cite{tangherlini}. These black holes have
support in asymptotically  AdS ($\Lambda<0$), dS ($\Lambda>0$), or
flat ($\Lambda=0$) backgrounds. Black hole solutions with
non-spherical topology (i.e., with $k=0$ and $k=-1$) live only in
an AdS background and do not have black hole counterparts in a
$\Lambda>0$ or in a $\Lambda=0$ background. The family with $k=0$
yields AdS black holes with planar, cylindrical or toroidal (with
genus $g= 1$) topology \cite{Toroidal_4D,Birmingham-TopDdim}).
Finally, the family with $k=-1$ yields AdS black holes with
hyperbolic, cylindrical, or toroidal topology with genus $g\geq 2$
\cite{topological,Birmingham-TopDdim,DehghaniRotTopol}). The
radial electromagnetic field produced by an electric charge
proportional to $Q$ is given by
\begin{equation}
F_{\mu\nu}=-\frac12\,\sqrt{\frac{(D-3)(D-2)}{2}}\,\frac{Q}{r^{D-2}}\,
\delta^0_\mu\,\delta^r_\nu\:. \label{maxwell}
\end{equation}

An important quantity in the matters discussed here is the surface
gravity $\kappa_{\rm h}$ defined at the horizon $r_h$ by
$\kappa_{\rm
h}^{2}=-\frac{1}{2}(\nabla^{\mu}\xi^{\nu})(\nabla_{\mu}\xi_{\nu})
{\bigl |}_{r_h}$, where $\xi^{\nu}$ is the timelike Killing vector
$\xi^{\nu}=(1,0,...,0)$. Using the metric (\ref{gen
metric})-(\ref{gen f}) one finds
\begin{equation}
\label{surfacegravity} \kappa_{\rm h}=-\frac12\, \frac{df(r)}{dr}
{\bigl |}_{r_h}\,.
\end{equation}
It is sometimes also useful to define the cosmological length
$\ell$ as
\begin{equation}
\label{cosm length} \ell^{2}=-\frac{3}{\Lambda}\,.
\end{equation}
For a detailed description of the properties of these black holes
see, e.g., \cite{DdimBH} and references therein.

%%%%%%%%%%%%%%%%%%%%%%%%%%%%%%%%%%%%%%%%%%%%%%%%%%%%%%%%%%%%%%%%%%%
\subsection{\label{sec:GEMS description}Brief description of the
GEMS procedure}
%%%%%%%%%%%%%%%%%%%%%%%%%%%%%%%%%%%%%%%%%%%%%%%%%%%%%%%%%%%%%%%%%%%%

In the next sections we will find the transformations
$z^{\alpha}(t,r,\theta_1,\cdots,\theta_{D-2})$ that perform a
global embedding of the above $D$-dimensional black hole
geometries into a $N$-dimensional Minkowskian spacetime (with
$N\geq D$) described by
\begin{equation}
\label{GEMS} ds^{2}=\eta_{\alpha \beta} dz^{\alpha}dz^{\beta},
\end{equation}
where $\eta_{\alpha \beta}$ is the flat metric in $N$-dimensions
with one or more timelike coordinates, and $\alpha,\beta =
0,1,2,...,N-1$. The GEMS procedure consists of these embedding
transformations. These transformations will map a Hawking
detector, that moves according to
\begin{equation}
\label{accelschw} r={\rm constant}\;, \;\;\theta_i={\rm
constant}\,\,,
\end{equation}
in the black hole geometry into an Unruh detector in the
$N$-dimensional flat spacetime that follows a hyperbolic
trajectory or Rindler motion described by
\begin{equation}
\label{accel} (z^{1})^{2}-(z^{0})^{2}=a^{-2},
\end{equation}
where $a$ is the constant acceleration of the Unruh detector in
the $N$-dimensional flat spacetime. With this GEMS procedure we
can compare the temperature (defined by Eq.(\ref{hawk})) measured
by the Hawking detector in the black hole spacetime with the
temperature (defined by Eq.(\ref{unruh})) measured by the Unruh
detector in the flat background.

The idea of the GEMS transformations is the following: First, the
line element containing the $(t,r)$ pair of coordinates is
transformed via a Rindler type transformation into a Minkowski
line element comprising the $(z^0,z^1)$ pair of coordinates. In
the process some terms containing $dr^2$ are left behind. Second,
one has to transform the angular part of the line element into a
flat space line element, through the usual procedure. In the
process some other terms containing $dr^2$ are again left behind.
Third, one now has to put these leftover $dr^2$ terms into a flat
form. Depending on the particular case, this can take zero, one or
two additional extra flat dimensions.

We first illustrate this procedure, with the simplest, trivial,
example: the $D$-dimensional Rindler geometry. In this case there
are no leftover $dr^2$ terms and thus this is the case where there
are zero additional extra flat dimensions. The metric is
\begin{equation}
\label{rin:metric} ds^{2}=-x^{2}\,\left(\frac{dt}{2}\right)^{2} +
dx^{2} + \sum_{i=2}^{D-1} (dx^i)^2\,,
\end{equation}
which has an acceleration horizon at $x=0$, (we use $t/2$ instead
of $t$ for convenience and consistency with what follows; see
Sec. \ref{sub:flat}).  This
Rindler geometry describes the metric seen by an uniformly
accelerated observer in a flat background: an observer at constant
$x$ and $x^i$'s suffers an acceleration $a=x^{-1}$. It is just an
wedge of Minkowski spacetime written in accelerated coordinates.
Consider now the timelike Killing vector $\xi^{\nu}=(1,0,...,0)$.
The surface gravity at the horizon $x=0$ is by definition, $k_{\rm
h}^{2}=-\frac{1}{2}(\nabla^{\mu}\xi^{\nu})(\nabla_{\mu}\xi_{\nu})
{\bigl |}_{x=0}$, which in the present case gives $k_{\rm h}=1$.
The Hawking temperature measured by a detector at constant $x$ and
$x^i$'s is then given by Eq.(\ref{hawk}):
\begin{equation}
\label{rin:temph} T_{\rm H}= \frac{x^{-1}}{2\pi}\,.
\end{equation}
Now, we can define the embedding of this $D$-dimensional Rindler
geometry in the $N$-dimensional Minkowski spacetime given by
Eq.(\ref{GEMS}). This is a trivial embedding since $N=D$:
\begin{eqnarray}
\label{rindler:gems}
z^{0} & = & x \sinh{\left(\frac{t}{2}\right)}\,, \nonumber\\
z^{1} & = & x \cosh{\left(\frac{t}{2}\right)}\,, \nonumber\\
z^{i} & = & x^{i}\,, \quad {\rm for} \:\:\: 2 \leq i \leq D-1\,.
\end{eqnarray}
The acceleration horizon $x=0$ is mapped into the plane $z^0=z^1$,
and transformations given by Eq.(\ref{rindler:gems}) constitute a
global embedding since they extend the original spacetime behind
the horizon and, in particular, they are analytic at the horizon.
Moreover, a Hawking detector that in the original Rindler
geometry, described by Eq.(\ref{rin:metric}), moved with constant
$x$ and $x^i$s is mapped through Eq.(\ref{rindler:gems}) into an
Unruh detector that moves along the hyperbolic trajectory of
Eq.(\ref{accel}), with uniform acceleration $a=x^{-1}$. This Unruh
detector measures a temperature given by Eq.(\ref{unruh}):
\begin{equation}
\label{rin:temph2} T_{\rm U}= \frac{x^{-1}}{2\pi}\,.
\end{equation}
Thus, the  Hawking temperature and the Unruh temperature match,
$T_{\rm H}=T_{\rm U}$.

In the next sections we will find the GEMS transformations for the
black holes presented in Sec. \ref{sec:BH Ddim}. These
transformations will be analytic at the horizon whose temperature
we will be computing.

%%%%%%%%%%%%%%%%%%%%%%%%%%%%%%%%%%%%%%%%%%%%%%%%%%%%%%%%%%%%%%%%%%%%
\section{\label{sec:curved}Global embedding of black holes in a
generic cosmological constant $\bm \Lambda$ background }
%%%%%%%%%%%%%%%%%%%%%%%%%%%%%%%%%%%%%%%%%%%%%%%%%%%%%%%%%%%%%%%%%%%%

In Sec. \ref{sub:menor} we give the global embeddings and the
matching of the Hawking and Unruh temperatures for higher
dimensional black holes in an asymptotically AdS background. Then
in sections \ref{sub:maior} and \ref{sub:flat} we study the
higher dimensional black holes in asymptotically dS and flat
backgrounds, respectively.

\subsection{\label{sub:menor}GEMS for the asymptotically AdS
black holes $\bm \Lambda<0$}
%%%%%%%%%%%%%%%%%%%%%%%%%%%%%%%%%%%%%%%%%%%%%%%%%%%%%%%%%%%%%%%%%%%%

The Einstein-Maxwell equations with a negative cosmological
constant ($\Lambda<0$) allow a family of static black hole
solutions, parameterized by a constant $k$, which can take the
values $k=1$ (horizons with spherical topology
\cite{tangherlini}), $k=0$ (planar, cylindrical or toroidal
topology \cite{Toroidal_4D,Birmingham-TopDdim}), and $k=-1$
(hyperbolic, cylindrical or toroidal topology
\cite{topological,Birmingham-TopDdim,DehghaniRotTopol}).

In the cases  $k=1,-1$ the $D$-dimensional black hole can be
globally embedded into a Minkowski background with $N=D+3$
dimensions, and in the case $k=0$ in a Minkowski background with
$N=2\times D$ dimensions. Note that the global embedding of  $D=4$
black holes with spherical topology ($k=1$) has been found in
\cite{deser3,kps}. The global embedding of black holes with
non-spherical topology ($k=0$, and $k=-1$) has not been discussed
at all, even for $D=4$.

For all three cases ($k=1,0,-1$) the first two dimensions $z^{0}$
and $z^{1}$ are given by the following GEMS transformations
\begin{eqnarray}
\label{maincoord} z^{0} & = & \kappa^{-1}_{\rm h}\sqrt{f(r)}
\sinh{(k_{\rm h}t)} \,, \nonumber\\
z^{1} & = & \kappa^{-1}_{\rm h}\sqrt{f(r)} \cosh{(k_{\rm h}t)} \,,
\end{eqnarray}
where $\kappa_{\rm h}$ is defined in (\ref{surfacegravity}).

For the $z^i$ coordinates related to the angular part of the
original line element, one has to separate the embeddings for  the
$k=1,-1$ cases from the embedding for $k=0$. For $k=1,-1$ one has
\begin{eqnarray}
\label{angularK} z^{2} & = & r \,S(\theta_{1})
\prod^{D-2}_{i=2}\sin(\theta_{i})
\,, \nonumber\\
z^{j} & = & r \,S(\theta_{1})
\bigg(\prod^{D-j}_{i=2}\sin(\theta_{i})\bigg)
\cos(\theta_{D+1-j})\,, \nonumber\\&& \hspace{4cm} {\rm for}
\:\: 3\leq j \leq D-1 \,, \nonumber\\
z^{D} & = & r\,C(\theta_{1})\,. \nonumber\\
\end{eqnarray}
where
\begin{eqnarray}
\label{S} S(\theta_{1})= \left\{\begin{aligned}
\sin{(\theta_{1})}&,\;\; \rm{ for }\;\; {\it k}=1 \\
\sinh{(\theta_{1})}&,\;\; \rm{ for }\;\;{\it k}=-1\\
\end{aligned}\right.\nonumber\\
\label{C} C(\theta_{1})= \left\{\begin{aligned}
\cos{(\theta_{1})}&,\;\; \rm{ for }\;\; {\it k}=1 \\
\cosh{(\theta_{1})}&,\;\; \rm{ for }\;\; {\it k}=-1 \\
\end{aligned}\right.
\end{eqnarray}
For $k=0$ one has
\begin{eqnarray}
\label{angularK2} z^{2j} & = & r \sin{\theta_{j}} \,, \hspace{1cm}
{\rm for} \:\: 1\leq j \leq D-2 \,,
\nonumber\\
z^{2j+1} & = & r \cos{\theta_{j}} \,, \hspace{1cm} {\rm for} \:\:
1\leq j \leq D-2 \,.
\end{eqnarray}
Note that the number of $z^i$ coordinates describing the angular
part in the $k=0$ cases is twice the number of angular coordinates
$\theta_i$ the black hole has.

The additional extra flat dimensions coming from the leftover
$dr^2$ terms can now be dealt with. The gravitational metric of
the higher dimensional charged AdS black holes is given by
Eq.(\ref{gen metric}) and Eq.(\ref{gen f}) with $\Lambda<0$. For
the three allowed values of the topological parameter $k$, the
corresponding black holes have two horizons, $r=r_-$ and $r=r_{\rm
h}$ (say) with $r_- \leq r_{\rm h}$, and $t$ is a timelike
coordinate for $r>r_{\rm h}$ \cite{DdimBH}. This is then the
region where a physical Hawking detector might be located, and it
is aware only of the $r=r_{\rm h}$ horizon. In particular, for our
purposes, the global embedding of this black hole into a flat
background will only have to cover the range $r>r_-$, that
includes a vicinity of $r=r_{\rm h}$. It is useful to swap the
mass parameter $M$ for the horizon radius $r_{\rm h}$. This is
achieved through the zero of  (\ref{gen f}), i.e.,
\begin{equation}
\label{chargeless:rh} M= k r_{\rm h}^{D-3}+\frac{Q^{2}}{r_{\rm
h}^{D-3}} +\frac{r_{\rm h}^{D-1}}{\ell^{2}}\,.
\end{equation}
In turn this allows us to write the function $f(r)$ in
Eq.(\ref{gen f}) as
\begin{eqnarray}
\label{charged:Frh}  f(r)&=& \frac{\ell^{2}(k\, r_{\rm h}^{D-3}\,
r^{D-3}-Q^{2})(r^{D-3}-r_{\rm
h}^{D-3})}{\ell^{2}\,r_{\rm h}^{D-3} \,r^{2(D-3)}} \nonumber \\
& &+ \frac{r_{\rm h}^{D-3} \, r^{D-3} (r^{D-1}-r_{\rm
h}^{D-1})}{\ell^{2}\,r_{\rm h}^{D-3} \,r^{2(D-3)}}\,.
\end{eqnarray}
The surface gravity associated with $r=r_{\rm h}$ is then
\begin{equation}
\label{charged:FL} \kappa_{\rm h} =  \frac { (D-3)\,\ell^{2}\,(k
\, r_{\rm h}^{2(D-3)}-Q^{2})+(D-1)\,r_{\rm
h}^{2(D-2)}}{2\,\ell^{2}\,r_{\rm h}^{2D-5}}\,.
\end{equation}
We are now ready to give the remaining GEMS coordinate
transformations that yields the additional extra flat dimensions.
The GEMS transformations for the case $k=1,-1$ are then
{\setlength\arraycolsep{2pt}
\begin{widetext}
\begin{eqnarray}
\label{charged:zd} z^{m_k}&=&\int dr \bigg[
\frac{W_{(D-3)}}{G(r)r^{D-3}}
+\frac{(D-3)^{2}\,W_{(D-1)}\big[\ell^{2}(k\, r_{\rm
h}^{2(D-3)}-Q^{2})+r_{\rm h}^{2(D-2)}\big]^{2}}{4
\,\ell^{4}\,k_{\rm h}^{2} \,G(r) \, r_{\rm h}^{3(D-3)+2}
\,r^{2(D-2)}} +\frac{Q^{2}(D-3)^{2}W_{(D-1)}}{\ell^{2}\,k_{\rm
h}^{2}\,G(r)\,r_{\rm h}^{D-3}\,r^{2(D-2)} }+H(k)
\bigg]^{\frac{1}{2}} \,, \nonumber \\
z^{n_k}&=&\int dr \bigg[ \frac{W_{(D-1)}\,r_{\rm h}^{D-3}\,r^{D-3}
+(D-3)(D-1)\ell^{2}Q^{2}W_{(D-3)}} {\ell^{4}\,k_{\rm h}^{2}\, G(r)
\,r_{\rm h}^{D-3}\,r^{2(D-3)}} +\frac{Q^{2}(D-3)^{2}}{k_{\rm
h}^{2}\, r^{2(D-2)}} +T(k) \bigg]^{\frac{1}{2}} \,,
\end{eqnarray}
\end{widetext}}
\noindent where $m_k$ and $n_k$ are defined by
\begin{eqnarray}
\label{um} {\rm for}\;\; k&=&1,-1 \left\{\begin{aligned}
m_k&\equiv& D+1 \\
n_k&\equiv& D+2\\
\end{aligned}\right.\nonumber\\
\label{zero} {\rm for}\;\; k&=&0\;\;\;\;\;\; \left\{\begin{aligned}
m_k&\equiv& 2D-2 \\
n_k&\equiv& 2D-1\\
\end{aligned}\right.
\end{eqnarray}
and the functions $H(k)$ and $T(k)$ are
\begin{eqnarray}
\label{Hterm}H(k)&\equiv&\delta_{-1k}\\
\label{Tterm}T(k)&\equiv&\delta_{1k}+\delta_{0k}(D-2)\,,
\end{eqnarray}
where $\delta_{\mu\nu}$ is the Kronecker symbol.
In addition we have defined $W_{(n)}$ as
\begin{equation}
\label{somatorio:rn} W_{(n)} \equiv \sum_{i=1}^{n}r_{\rm
h}^{i-1}r^{n-i}\,.
\end{equation}
and $G(r)$ as
\begin{eqnarray}
\label{charged:func}& & \hspace{-0.5cm} G(r)= f(r)/(r-r_{\rm h}) \nonumber \\
&= & \frac {\ell^{2}(k\,r_{\rm h}^{D-3}\,r^{D-3}-Q^{2})W_{(D-3)}
+r_{\rm h}^{D-3}W_{(D-1)}\,r^{D-3}}
{\ell^{2}\,r_{\rm h}^{D-3}\,r^{2(D-3)}}\,. \nonumber \\
\end{eqnarray}
A careful analysis of the arguments of the square roots in
Eq.(\ref{charged:zd}) shows that they are always positive and
finite for $r>r_-$ and, in particular, these GEMS transformations
are analytical at $r=r_{\rm h}$.

Thus, a spherical $(k=1)$ $D$ dimensional charged black hole in an
AdS background can be embedded in a  $D+3$ dimensional Minkowski
spacetime with two times, with metric
$ds^{2}=-(dz^{0})^{2}+\sum_{i=1}^{D+1}(dz^{i})^{2}-(dz^{D+2})^{2}$.
A hyperbolic $(k=-1)$ $D$ dimensional charged black hole in an AdS
background can be embedded in a $D+3$ dimensional Minkowski
spacetime with three times, with metric
$ds^{2}=-(dz^{0})^{2}+\sum_{i=1}^{D-1}(dz^{i})^{2}-(dz^{D})^{2}+
(dz^{D+1})^{2}-(dz^{D+2})^{2}$. Finally, a planar $(k=0)$ $D$
dimensional charged black hole in an AdS background can be
embedded in a $2\times D$ dimensional Minkowski spacetime with two
times, with metric
$ds^{2}=-(dz^{0})^{2}+\sum_{i=1}^{2D-2}(dz^{i})^{2}-(dz^{2D-1})^{2}$.

These GEMS transformations map the horizon at $r=r_{\rm h}$ into
the plane $z^1=z^0$, and the original Hawking detector is mapped
into the Unruh detector that follows the hyperbolic motion
described by Eq.(\ref{accel}), with constant acceleration
$a=r_{\rm h}/\sqrt{f(r)}$. This Unruh detector then measures a
temperature given by Eq.(\ref{unruh}) which matches with the
temperature given by Eq.(\ref{hawk}) measured by its Hawking
partner in the original black hole. As a consistency check we
remark that the GEMS transformations given in
Eqs.(\ref{maincoord})-(\ref{charged:func}) for the
Reissner-Nordstr\"om ($D=4$ and $k=1$) black hole reduce to the
transformations found in \cite{kps}.

The gravitational metric of the higher dimensional $k=1,0,-1$
neutral AdS black holes is given by Eq.(\ref{gen metric}) and
Eq.(\ref{gen f}) with $Q=0$ and $\Lambda<0$. This case is in
everything similar to the charged black hole. To find the GEMS one
just has to make the limit $Q=0$ in the
Eqs.(\ref{maincoord})-(\ref{charged:func}). As a consistency check
we remark that the GEMS transformations for the neutral
Schwarzschild ($D=4$ and $k=1$) black hole reduce to the
transformations found in \cite{deser3}. And when we further set
$M=0$ we recover the embedding of the pure AdS spacetime found in
\cite{deser3}.

Since extremal black holes, ($r_{\rm h}=r_{-}$), have zero
temperature, there is no interest here in discussing their GEMS
transformations. Indeed, the GEMS transformations showed above do
not even apply to extremal cases.

%%%%%%%%%%%%%%%%%%%%%%%%%%%%%%%%%%%%%%%%%%%%%%%%%%%%%%%%%%%%%%%%%%%
\subsection{\label{sub:maior}GEMS for the asymptotically dS black
holes $\bm \Lambda>0$}
%%%%%%%%%%%%%%%%%%%%%%%%%%%%%%%%%%%%%%%%%%%%%%%%%%%%%%%%%%%%%%%%%%%

When $\Lambda>0$, the Einstein-Maxwell equations allow black hole
solutions only for $k=1$, i.e., whose horizons have a spherical
topology, see Eqs.(\ref{gen metric})-(\ref{maxwell}).

In the charged case, $Q\neq 0$, the solution has three horizons,
the Cauchy horizon $r=r_-$, the event horizon $r=r_{\rm h_+}$, and
the cosmological horizon $r=r_{\rm h_c}$, with $r_-\leq r_{\rm
h_+} \leq r_{\rm h_c}$. A physical detector can be in between
$r_{\rm h_+}$ and $r_{\rm h_c}$, and it is aware only of $r_{\rm
h_+}$ and $r_{\rm h_c}$. Use of
Eqs.(\ref{maincoord})-(\ref{charged:func}) with  $\ell^2
\rightarrow -\ell^2$ and $r_{\rm h}\equiv r_{\rm h_+}$ yields an
embedding that covers  the region $r_-<r<r_{\rm h_c}$. This
embedding allows us to verify the matching between the Hawking
temperature and the Unruh temperature of the black hole horizon
$r=r_{\rm h_+}$. Alternatively, if we instead use
Eqs.(\ref{maincoord})-(\ref{charged:func}) with $\ell^2
\rightarrow -\ell^2$ and $r_{\rm h}\equiv r_{\rm h_c}$, one gets
an embedding that covers the region $r>r_{\rm h_+}$ and comprises
the cosmological horizon. We can then use this embedding to verify
that $T_{\rm H}=T_{\rm U}$ also for the cosmological horizon.

In the neutral case, $Q=0$, the black hole has no Cauchy horizon,
and the discussion of the last paragraph follows by replacing
$r_-$ by $r_-=0$.

%%%%%%%%%%%%%%%%%%%%%%%%%%%%%%%%%%%%%%%%%%%%%%%%%%%%%%%%%%%%%%%%%%%
\subsection{\label{sub:flat}GEMS for the asymptotically
flat black holes $\bm \Lambda=0$}
%%%%%%%%%%%%%%%%%%%%%%%%%%%%%%%%%%%%%%%%%%%%%%%%%%%%%%%%%%%%%%%%%%%

The charged Tangherlini black hole, which is the higher
dimensional Reissner-Nordstr\"{o}m black hole, has been found in
\cite{tangherlini}. Its gravitational field is described by
Eq.(\ref{gen metric}) and Eq.(\ref{gen f}) with $k=1$ and
$\Lambda=0$, and its Maxwell field is given by Eq.(\ref{maxwell}).
It has two horizons, the event horizon at $r^{D-3}_{\rm h} =M/2+
\sqrt{M^{2}/4-Q^{2}}$, and the Cauchy horizon at $r^{D-3}_{-}
=M/2- \sqrt{M^{2}/4-Q^{2}}$ with $r_-\leq r_{\rm h}$. In
particular, for our purposes, the global embedding of this black
hole into a flat background will only have to cover the region
$r>r_-$, that includes a vicinity of both sides of $r=r_+$.

The GEMS transformations are given by
Eqs.(\ref{maincoord})-(\ref{charged:func}) with $k=1$ and
$\Lambda=0$. In order to make touch with the results of
\cite{deser3} we will give explicitly for this case the
transformations of the $z^{D+1},z^{D+2}$ coordinates in terms of
the horizon radial coordinates $r_h$ and $r_-$. From
Eq.(\ref{charged:zd}) one finds {\setlength\arraycolsep{2pt}
\begin{widetext}
\begin{eqnarray}
\label{rn:gems} z^{D+1} &=& \int dr  \bigg(\frac
{[W_{(D-1)}-(r_{\rm h}^{D-3}r+r_{\rm h}^{D-2})](r_{\rm h}^{D-3}
+r_{-}^{D-3})+r r_{\rm h}^{2(D-3)} +r_{\rm h}^{2D-5}}
{r^{2}(r^{D-3}-r_{-}^{D-3})
W_{(D-3)}}\bigg)^{\frac{1}{2}} \,, \nonumber\\
z^{D+2} & = & \int dr \bigg(\frac{4r_{\rm h}^{3D-7}
r_{-}^{D-3}}{(r_{\rm h}^{D-3}-r_{-}^{D-3})^{2}\,r^{2(D-2)}}\bigg)
^{\frac{1}{2}} \,.
\end{eqnarray}
\end{widetext}}
\noindent When $D=4$ these transformations reduce to those found
in \cite{deser3}. These transformations map the horizon at
$r=r_{\rm h}$ into the plane $z^1=z^0$, and are complete due to
the analyticity of $z^{D+1}$ and $z^{D+2}$ in the range $r>r_-$.
Note also that the surface gravity associated with the $r=r_{\rm
h}$ horizon is $k_{\rm h}=\frac{D-3}{2} (r_{\rm
h}^{D-3}-r_{-}^{D-3})/r_{\rm h}^{D-2}$ and its temperature
measured by a Hawking detector at constant $r$ and $\theta_i$'s,
is given by Eq.(\ref{hawk}), which matches the Unruh temperature
(\ref{unruh}).

The higher dimensional Schwarzschild black hole, the Tangherlini
black hole, is found by setting $Q=0$ in the higher dimensional
Reissner-Nordstr\"om black hole. Its gravitational field is
described by Eq.(\ref{gen metric}) and Eq.(\ref{gen f}) with
$k=1$, $Q=0$ and $\Lambda=0$. In the $Q=0$ limit, one has $r_-=0$
and $z^{D+2}=0$ in Eq.(\ref{rn:gems}), and this limit yields the
Schwarzschild GEMS. Note that in the limit
$M\rightarrow \infty$, the near horizon geometry is the Rindler
metric and one easily recovers
Eqs. (\ref{rin:metric})-(\ref{rin:temph2}).

%%%%%%%%%%%%%%%%%%%%%%%%%%%%%%%%%%%%%%%%%%%%%%%%%%%%%%%%%%%%%%%%%%%
\section{\label{sec:conclusion}Conclusion}
%%%%%%%%%%%%%%%%%%%%%%%%%%%%%%%%%%%%%%%%%%%%%%%%%%%%%%%%%%%%%%%%%%%

We have discussed the thermal properties of higher dimensional
black holes. We have verified that the Hawking temperature matches
the Unruh temperature, by mapping the curved space observers into
the Rindler-accelerated observers of a higher dimensional
spacetime, a formulation introduced by Deser and Levin
\cite{deser3} in the $D=4$ case. In order to accomplish this task
we have found the global embedding Minkowskian spacetime (GEMS)
transformations from $D$-dimensional black holes into a higher
dimensional Minkowskian spacetime. The global embedding of AdS black
holes with non-spherical topology ($k=0$, and $k=-1$) had not
been discussed previously, even for $D=4$. The GEMS transformations are
important on their own, since they provide a powerful tool that
simplifies the study of black hole physics by working instead, but
equivalently, in a flat background geometry.

\begin{acknowledgments}
This work was partially funded by Funda\c c\~ao para a Ci\^encia e
Tecnologia (FCT) -- Portugal through project CERN/FNU/43797/2001.
N.L.S. acknowledges financial support from FCT through grant
SFRH/BD/2003. O.J.C.D. acknowledges financial support from FCT
through grant SFRH/BPD/2003. J.P.S.L. thanks Observat\'orio
Nacional do Rio de Janeiro for hospitality.
\end{acknowledgments}

\end{document}